\documentclass[english]{article}
\usepackage[latin9]{inputenc}
\usepackage{amsmath}
\usepackage{amssymb}
\usepackage{esint}

\makeatletter

\usepackage{amsthm}\usepackage[english]{babel}

\newcommand{\tr}{\mathop{\rm tr\,}\nolimits}
\newtheorem{lemma}{Lemma}
\newtheorem{theorem}{Theorem}

\date{}

\title{Linear Hamiltonian Systems under Microscopic Random Influence}

\author{Lykov A. A. \thanks{Lomonosov Moscow State University, Faculty of Mechanics and Mathematics, Moscow, Russia; e-mail: alekslyk@yandex.ru, maly\-shev2@yahoo.com, stepan\_muzychka@mail.ru}\and Malyshev V. A.\footnotemark[1] \and Muzychka S. A. \footnotemark[1]}

\makeatother

\usepackage{babel}
\begin{document}
\maketitle
\begin{abstract}
It is known that a linear hamiltonian system has too many invariant
measures, thus the problem of convergence to Gibbs measure has no
sense. We consider linear hamiltonian systems of arbitrary finite
dimension and prove that, under the condition that one distinguished
coordinate is subjected to dissipation and white noise, then, for
almost any hamiltonians and almost any initial conditions, there exists
the unique limiting distribution. Moreover, this distribution is Gibbsian
with the temperature depending on the dissipation and of the variance
of the white noise. 
\end{abstract}

\setcounter{page}{1}

\section{Main Results}

\ Consider the phase space 
\[
L=L_{2N}=\mathbb{R}^{2N}=\{\psi=(q,p):\ q=(q_{1},\ldots,q_{N}),\ p=(p_{1},\ldots,p_{N})\in\mathbb{R}^{N}\}
\]
with the scalar product 
\[
(\psi,\psi')_{1}=\sum_{i=1}^{N}(q_{i}q_{i}'+p_{i}p_{i}').
\]
The space $L$ is the direct sum $L=l_{N}^{(q)}\oplus l_{N}^{(p)}$
of orthogonal coordinate space and momentum space with the induced
scalar products $(q,q')_{1}$ and $(p,p')_{1}$ correspondingly. We
are most interested in the case of large $N$, but we do not use it
in the present paper.

We shall study the following system of $2N$ stochastic differential
equations ($k=1,\ldots,N$) 
\begin{equation}
\begin{array}{rcl}
dq_{k} & = & p_{k}\, dt,\\[4pt]
dp_{k} & = & \sum_{l=1}^{N}\Big((-V(k,l)q_{l}-D(k,l)p_{l})\, dt+B(k,l)\, dw_{t,l}\Big),
\end{array}\label{system_2N}
\end{equation}
where $V=(V(i,j))$ is a positive definite $(N\times N)$-matrix,
$D=(D(i,j))$ is the non-negative definite symmetric $(N\times N)$-matrix,
$B=(B(k,l))$ is an arbitrary real matrix, $w_{t,l}$, $l=1,\ldots,N,$
are the standard brownian processes, independent in $l$.

If $D=B=0$, then the system is a linear hamiltonian system with the
quadratic hamiltonian 
\begin{equation}
H(\psi)=T+U,\qquad T=\frac{1}{2}\sum_{i=1}^{N}p_{i}^{2},\quad U=\frac{1}{2}\sum_{i,j=1}^{N}V(i,j)q_{i}q_{j}.\label{hamiltonian}
\end{equation}

We will consider systems where the matrices $B$ and~$D$ are as
follows (for some $n=1,\ldots,N$) 
\[
D=\alpha\Delta_{n},\quad B=\sigma\Delta_{n},
\]
where $\Delta_{n}=(\delta_{i,j}\delta_{i,n})$ is the matrix with
one diagonal element equal to $1$ and other elements are all zero.
Note that in this case the system will be subjected to only one white
noise $dw_{t,n}$. We assume that the index $n$ is fixed, and write
for shortness $dw_{t}=dw_{t,n}$. Further on, unless otherwise stated,
we assume that $\alpha>0$, $\sigma>0$.

System~(\ref{system_2N}) can be rewritten in the matrix form 
\begin{equation}
d\psi=A\psi\, dt+\sigma g_{n}\, dw_{t},\label{main_matrix_form}
\end{equation}
where 
\[
A=\left(\begin{array}{cc}
0 & \quad E\\
-V & \quad-D
\end{array}\right),
\]
$E$ is the unit $(N\times N)$-matrix, $g_{n}=(0,e_{n})\in l_{N}^{(p)}$,
and $e_{n}=(0,\ldots,0,1,0,\ldots,0)\in\mathbb{R}^{N}$ is\ $n$-th
standard basis vector. The solution of the latter equation with arbitrary
initial vector $\psi(0)$ is uniquely defined and can be written as
(for example, see section 12.4 of \cite{Ventsel}) 
\[
\psi(t)=e^{tA}\bigg(\sigma\int_{0}^{t}e^{-sA}g_{n}\, dw_{s}+\psi(0)\bigg).
\]
Introduce the set 
\[
L_{-}=\{\psi\in L:\ H(e^{tA}\psi)\to0,\ t\to\infty\}\subset L.
\]
We will need the following results.

\begin{lemma} The set $L_{-}$ is a linear subspace of the space
$L,$ and moreover $L_{-}=\{(q,p)\in L:\ q\in l_{V},\ p\in l_{V}\}$
where $l_{V}$~--is the subspace of $\ \mathbb{R}^{N},$ generated
by the vectors $V^{k}e_{n},\ k=0,1,\ldots,$ in particular $g_{n}\in L_{-}$.
Moreover, $L_{-}$ and its orthogonal complement, denoted further
by $L_{0},$ are invariant with respect to the operator $A$. \end{lemma}

All assertions of this lemma have been proven in \cite{LM_1}.

By this lemma any initial vector $\psi(0)$ can be uniquely decomposed
as 
\[
\psi(0)=\psi_{0}+\psi_{-},\qquad\psi_{0}\in L_{0},\quad\psi_{-}\in L_{-}.
\]
Then the solution $\psi(t)$ of the stochastic equation~(\ref{main_matrix_form})
with initial vector $\psi(0)$ for any $t\geq0$ can be decomposed
as 
\begin{equation}
\psi(t)=\psi^{(0)}(t)+\psi^{(-)}(t),\label{expansion}
\end{equation}
where $\psi^{(0)}(t)$, $\psi^{(-)}(t)$ satisfy the equations 
\[
\dot{\psi}^{(0)}(t)=A\psi^{(0)}(t),\quad d\psi^{(-)}(t)=A\psi^{(-)}\, dt+\sigma g_{n}\, dw_{t}
\]
with the initial conditions $\psi^{(0)}(0)=\psi_{0}$, $\psi^{(-)}(0)=\psi_{-}$
correspondingly. In fact, the sum of these equations give the equation~(\ref{main_matrix_form}).

By Lemma 1, the function $\psi^{(0)}(t)\in L_{0}$, $t\in[0,\infty),$
is deterministic and can be written as 
\begin{equation}
\psi^{(0)}(t)=e^{At}\psi^{(0)}(0),\label{psi_zero}
\end{equation}
and $\psi^{(-)}(t)$ is a gaussian random process with values in $L_{-}$
(as $g_{n}\in L_{-}$).

\begin{theorem} For any $\psi(0)$ the convergence in distribution
\[
\psi^{(-)}(t)\mathop{\longrightarrow}\limits _{t\to\infty}\xi,
\]
takes place, where $\xi\in L_{-},$ and its distribution is absolutely
continuous with respect to Lebesgue measure on $L_{-}$ (defined by
the euclidean structure), and has the following density with respect
to this measure 
\begin{equation}
p_{\xi}(\psi)=\frac{1}{Z}\exp\bigg(-\frac{2\alpha}{\sigma^{2}}\, H(\psi)\bigg),\qquad\psi\in L_{-}.\label{Gibbs_density}
\end{equation}

The limit of the mean energy is: 
\[
\lim_{t\rightarrow+\infty}\bold{E}\, H(\psi^{(-)}(t))=\frac{\sigma^{2}}{4\alpha}\dim{L_{-}}.
\]
\end{theorem}

Thus, the action of the random force and dissipation on one particle
only garanties convergnce to the invariant Gibbs measure with the
temperature depending on $\alpha$ and $\sigma$.

The first assertion of the next theorem shows that convergence to
Gibbs distribution is a typical property of linear hamiltonian systems
with dissipation and random force. The second assertion shows that
the dissipative term is necessary for this convergence.

For given $N$ denote $\mathbf{H}_{N}$ the smooth manifold of all
possible hamiltonians $H$ as in (\ref{hamiltonian}), that is \ the
smooth manifold of all positive definite $(N\times N)$-matrices $V$.
Let $\mu$ be an arbitrary absolutely continuous probability measure
on $\mathbf{H}_{N}$, and let $\mathbf{H}_{N}^{(+)}$ be the set of
all hamiltonians of $\mathbf{H}_{N}$, for which the dimension of
$L_{0}$ is greater than zero.

\begin{theorem}\ 1.~The measure $\mu$ of the subset $\mathbf{H}_{N}^{(+)}$
is zero. 2.~If $\alpha=0,$ then for any initial condition $\psi(0)$
we have 
\[
\mathbf{E}\, H(\psi(t))=\frac{\sigma^{2}}{2}\, t+O(1).
\]
\end{theorem}

Note that for more restricted (physical) classes of hamiltonians the
property $\dim L_{0}=0$ is not typical (see \cite{LM_1} in this
respect).

In this short note we restrict ourselves to the most interesting case
of one distinguished particle, which shows that even the minimal introducing
of stochasticity to the system garanties the convergence to the physical
equilibrium. Note however, that most results can be generalized to
arbitrary matrices $D$ and $B$. Similar systems, mainly one-dimensional
(one of the goal was to justify the Fourier law of heat conduction)
were considered in the 1960-70 in the series of papers by J. Lebowitz
and colleagues (see~\cite{leb_rieder},~\cite{leb_spohn} and references
therein).

\section{Proofs}

\quad{}\\

\vspace*{-8pt}
 Proof of theorem 1. Assume first that $\dim L_{0}=0$. In this case
the spectrum of the matrix $A$ belongs to the left half-plane.

Let us prove the convergence first. By the latter assumption, we can
consider only the process $\psi^{(-)}(t)$, admitting the following
decomposition 
\[
\begin{array}{c}
\psi^{(-)}(t)=\psi^{(g)}(t)+\psi^{(d)}(t),\\[4pt]
\psi^{(g)}(t)=\sigma e^{tA}\int_{0}^{t}e^{-sA}g_{n}\, dw_{s},\quad\psi^{(d)}(t)=e^{tA}\psi^{(-)}(0).
\end{array}
\]
By definition of $L_{-}$ the function $\psi^{(d)}(t)$ tends to zero
if $t\rightarrow+\infty$. That is why it is sufficient to prove the
convergence of $\psi^{(g)}(t)$. Denote $C(t)=(\mathbf{E}\,\{\psi_{i}^{(g)}(t)\psi_{j}^{(g)}(t)\})$
the covariance matrix. Then 
\[
C(t)=\mathbf{E}\,\{\psi^{(g)}(t)(\psi^{(g)})^{T}(t)\}=\sigma^{2}e^{tA}\,\mathbf{E}\,\bigg\{\int_{0}^{t}e^{-sA}g_{n}\ \, dw_{s}\int_{0}^{t}g_{n}^{T}e^{-sA^{T}}\, dw_{s}\bigg\} e^{tA^{T}},
\]
where $^{T}$ denotes transposition. Using the Ito isometry \cite{BulShir},
we get 
\begin{equation}
C(t)=\sigma^{2}e^{tA}\int_{0}^{t}e^{-sA}g_{n}g_{n}^{T}e^{-sA^{T}}\, ds\, e^{tA^{T}}.\label{C_integral}
\end{equation}
Let us calculate the integral in the last formula by finding the matrix
$U$, not depending on time and such that 
\begin{equation}
\int_{0}^{t}e^{-sA}g_{n}g_{n}^{T}e^{-sA^{T}}\, ds=e^{-tA}Ue^{-tA^{T}}-U.\label{matrix_U}
\end{equation}
Differentiation (\ref{matrix_U}) in $t$ shows that the exponents
cancel and we get 
\[
AU+UA^{T}=-g_{n}g_{n}^{T}.
\]
This equation with respect to $U$ has the unique solution, as the
spectrum of $A$ lies in the left half-plane, see section 4.4 of \cite{DalKrein}.
It is easy to check that the solution is the following matrix 
\begin{equation}
U=\frac{1}{2\alpha}\left(\begin{array}{cc}
V^{-1} & \quad0\\
0 & \quad E
\end{array}\right).\label{matrix_U_explicit}
\end{equation}

Thus, from (\ref{C_integral}) and (\ref{matrix_U}) we get $C(t)=\sigma^{2}(U-e^{tA}Ue^{tA^{T}})$.
As the spectrum of $A$ lies in the left half-plane, then $\lim_{t\rightarrow+\infty}C(t)=\sigma^{2}U.$
That is why the following limit in distribution 
\[
\xi=\lim_{t\to\infty}\psi^{(g)}(t)=\lim_{t\to\infty}\psi^{(-)}(t),
\]
exists and is the gaussian vector with zero mean and covariance matrix
$\sigma^{2}U$.

Now prove the last assertion of theorem 1. From (\ref{matrix_U_explicit})
and from positive definiteness of $V$, it follows that $U$ is non-degenerate.
Thus the distribution of $\xi$ has the density $p_{\xi}(\psi)$ with
respect to the standard Lebesgue measure $dq_{1}\cdots dp_{N}$ on
$L$. As the matrix $(2\alpha U)^{-1}$ defines the quadratic form
$H$, then 
\[
p_{\xi}(\psi)=\frac{1}{Z}\exp\bigg(-\frac{2\alpha}{\sigma^{2}}\, H(\psi)\bigg).
\]
Remind that $H(\psi)=(Q\psi,\psi)_{1}/2$, where the $(2N\times2N)$-matrix
$Q=(q_{i,j})$ is defined by the equality of the vectors $Q(q,p)=(Vq,p)$.
Then for the mean energy we get 
\[
\lim_{t\rightarrow+\infty}\mathbf{E}\, H(\psi^{(-)}(t))=\mathbf{E}\, H(\xi)=\frac{1}{2}\sum_{i,j=1}^{2N}q_{i,j}c_{i,j}=\frac{\sigma^{2}}{2}\tr(QU)=\frac{\sigma^{2}}{4\alpha}\dim L.
\]
Thus for the case $L_{0}=\{0\}$ theorem 1 is proved.

Consider now the case $L_{0}\ne\{0\}$. Let $v_{1},\ldots,v_{d}$
be an arbitrary orthonormal basis of the subspace $l_{V}$ with the
only restriction that $v_{1}=e_{n}$. Using it, define the orthonormal
basis in $L_{-}$ as follows 
\[
h_{k}^{(q)}=(v_{k},0),\quad h_{k}^{(p)}=(0,v_{k}),\qquad k=1,\ldots,d.
\]
The coordinates on $L_{-}$ in this basis we denote $\psi'=(\psi_{1}',\ldots,\psi_{2d}')$.
In these coordinates our equation on $L_{-}$ can be written as 
\[
d\psi'=A'\psi'\, dt+g'_{1}\, dw_{t},
\]
where $A'=\footnotesize{\left(\begin{array}{cc}
0 & E\\
-V' & -D'
\end{array}\right)}$, $E$ is the unit matrix of the order $d$, $V'$, $D'$ and $g'_{1}$
are the matrices of the operators $V$, $D$ and the distinguished
vector, in the new coordinates, correspondingly.

The latter equation looks like the main equation~(\ref{main_matrix_form}).
The operator~$A'$ is the restriction of the operator~$A$ onto
the subspace $L_{-}$; let $L_{-}=L_{0}'\oplus L_{-}'$\ be the correponding
decomposition of $L_{-}$ in $A'$. Then $L'_{0}=0$ and we can apply
the assertions proven for the case $L_{0}=0$. Note that the quadratic
form $H'$ on $L_{-}$, generated by $V'$, coincides with the restriction
of the quadratic form ~$H$ on the subspace $L_{-}$. This proves
Theorem 1 completely.

Proof of theorem 2. If $\alpha=0$, then the energy conservation law
gives $L_{-}=0$. Decompose the solution $\psi(t)$ as 
\[
\psi(t)=\psi_{h}(t)+\psi_{I}(t),
\]
where $\psi_{h}(t)$-is the solution of the homogeneous equation with
initial condition $\psi(0)$, and $\psi_{I}(t)$ is the solution of
the inhomogeneous equation with zero initial conditions. As $H(\psi(t))$
has the norm properties, then 
\[
|H(\psi(t))-H(\psi_{I}(t))|\leqslant H(\psi_{h}(t))=H(\psi(0)).
\]
The latter equality holds as $\alpha=0$, and thus the energy is conserved.
Let us find $\mathbf{E}\, H(\psi_{I}(t))$. Denote $\psi_{I}(t)=(q^{(I)}(t),p^{(I)}(t))$.
The solution with zero initial conditions is similar to the known
(see for example \cite{Gant}) formula for ordinary differential equations
\begin{eqnarray*}
q(t) & = & \sigma\ (\sqrt{V})^{-1}\int_{0}^{t}\sin(\sqrt{V}(t-s))e_{n}\, dw_{s},\\
p(t) & = & \sigma\ \int_{0}^{t}\cos(\sqrt{V}(t-s))e_{n}\, dw_{s}.
\end{eqnarray*}
Using again the Ito isometry we get 
\begin{eqnarray*}
\mathbf{E}\, T & = & \frac{\sigma^{2}}{2}\,\mathbf{E}\,(p(t),p(t))_{1}=\frac{\sigma^{2}}{2}\,\mathbf{E}\,\{p^{T}p\}=\frac{\sigma^{2}}{2}\int_{0}^{t}e_{n}^{T}\cos^{2}\big(\sqrt{V}(t-s)\big)e_{n}\, ds,\\
\mathbf{E}\, U & = & \frac{\sigma^{2}}{2}\,\mathbf{E}\,(Vq(t),q(t))_{1}=\frac{\sigma^{2}}{2}\,\mathbf{E}\{q^{T}Vq\}=\frac{\sigma^{2}}{2}\int_{0}^{t}e_{n}^{T}\sin^{2}\big(\sqrt{V}(t-s)\big)e_{n}\, ds,\\
\mathbf{E}\, H(\psi_{I}(t)) & = & \mathbf{E}\,\{T+U\}=\frac{\sigma^{2}}{2}\int_{0}^{t}e_{n}^{T}e_{n}\, ds=\frac{\sigma^{2}}{2}\, t.
\end{eqnarray*}
We have thus proved the second assertion of theorem 2.

Note that $\mathbf{H}_{N}$ is a smooth manifold. Define the matrix
$\Sigma(V)$ so that its $k$-th column is equal to the vector $V^{k}e_{n}$.
Then 
\[
\mathbf{H}_{N}^{(+)}=\{V:\ \dim l_{V}<N\}=\big\{ V:\ \det(\Sigma(V))=0\big\}.
\]
Note that $\det(\Sigma(V))\ne0$ for matrices $V\in\mathbf{H}_{N}$
with simple spectrum and eigenvalue basis $v_{1},\ldots,v_{N}$, with
the property that $(v_{k},e_{n})_{1}\ne0$ for all $k=1,\ldots,N$,
see \cite{LM_1}. That is why the function $\det(\Sigma(V))$ is not
identically zero on the manifold $\mathbf{H}_{N}$. Thus, $\mathbf{H}_{N}^{(+)}$
is the set of zeros of the polynomial on $\mathbf{H}_{N}$. It follows
that its dimension is less than the dimension of all $\mathbf{H}_{N}$,
and its measure~$\mu$ equals zero.

\end{document}